# Current-field diagram of magnetic states of a surface spin valve in a point contact with a single ferromagnetic film

I.K. Yanson[1], O.P. Balkashin[1], V.V. Fisun[1], Yu.I. Yanson[2], and Yu.G. Naidyuk[1]

[1]*B. Verkin Institute for Low Temperature Physics and Engineering of the National Academy of Sciences of Ukraine*
*47 Lenin Ave., Kharkov 61103, Ukraine*
E-mail: naidyuk@ilt.kharkov.ua

[2]*Kamerlingh Onnes Laboratory, Leiden University, Niels Bohrweg 2, 2333 CA Leiden, The Netherlands*

We present a study of the influence of an external magnetic field $H$ and an electric current $I$ on the spin-valve (SV) effect between a ferromagnetic thin film (F) and a sharp tip of a nonmagnetic metal (N). To explain our observations, we propose a model of a local surface SV which is formed in such a N/F contact. In this model, a ferromagnetic cluster at the N/F interface plays the role of the free layer in this SV. This cluster exhibits a larger coercive field than the bulk of the ferromagnetic film, presumably due to its nanoscale nature. Finally, we construct a magnetic state diagram of the surface SV as a function of $I$ and $H$.

## 1. Introduction

Spin-valve (SV) effect on the electric conductance of $F_1/N/F_2$-type nanopillars, where $F_{1,2}$ represent ferromagnetic layers and N is a nonmagnetic spacer layer, generated great interest both from the fundamental point of view and from its application perspective in spintronic devices [1]. The fundamental interest lies in the spin-transfer torque (STT) in a SV by a spin-polarized current that passes through it [2]. This effect could be used to control the magnetic state of a SV [3]: the passage of a spin-polarized current can change the orientation of the magnetization **M** of the layers $F_1$ and $F_2$ relative to each other, depending on the direction of the current flow. Hence, if one would fix the magnetization of the layer $F_1$ (fixed layer), the magnetization of the other layer $F_2$ (free layer) could be set either parallel (P) or antiparallel (AP) to $F_1$ depending on the direction of an electric current that flows through the system. These two states, i.e., the P and the AP states, lead to a different conductivity of a SV due to the giant magnetoresistance effect [4]. This results in a hysteresis in the resistance of a SV with a transition between the low resistance state (P orientation) and a high resistance state (AP orientation) during a bipolar current sweep. In addition, the direction of $M_1$ and $M_2$ can also be controlled by an external magnetic field. If the layers $F_1$ and $F_2$ have different coercivities, the P and AP orientations of $M_{1,2}$ can be realized, forming two states in the magnetoresistance $R(H)$ of a SV. These states produce hysteresis loops (meanders) in the $R(H)$ dependence within the range of the coercive fields of the fixed and the free layers.

In particular, one is interested in the behavior of a SV structure both in an external magnetic field and under electric current. This situation has been studied in pillar-type SV structures based on Co and permalloy ($Ni_{80}Fe_{20}$), for which current-field ($I$–$H$) diagrams of the magnetic state of a SV were obtained [5–7]. The effects of a flow of electrons that creates a spin transfer torque, and a magnetic field **H** can mutually enhance or suppress each other, leading to an asymmetric $I$–$H$ phase diagram. Additionally, as proposed in Ref. 6, other factors, such as the self-Oersted-field of the current, generation of non-equilibrium magnons, and thermal effects due to heating of the structure by a high current density, may influence the shape of the $I$–$H$ phase diagram.

Spin-valve effects that are similar to those found in the lithographically-made conducting $F_1/N/F_2$ nanopillars were also observed in point contacts between a single ferromagnetic thin film and a nonmagnetic metallic tip [8–10]. According to Refs. 8, 9, the role of the free layer $F_2$ in such a system is taken by a ferromagnetic domain that is formed at the point contact. However, a study of this effect in Co thin films of varying thickness (3–100 nm) and point contacts with diameters from several tens down to few nanometers raised doubts about the formation of "conventional" ferromagnetic domains as an explanation of this effect [10]. Instead, a model of a surface SV (SSV) has been proposed, where a ferromagnetic layer of a few atoms thick at the surface, which has a weakened magnetic bond to the bulk of the film, plays the role of the free layer $F_2$.

Recently it was shown that other effects that are characteristic of the $F_1/N/F_2$ pillar structures can also be observed on a single ferromagnetic film [11–13]. These effects include the dynamic SV effect [11], formation of spin vortex states [12], and exchange bias effect [13]. The size of the point contacts in these studies was at least an order of magnitude smaller than the existing lithographically-prepared $F_1/N/F_2$ pillars, thus showing that the SV effect is preserved on the nanoscale.

In this paper we present our study on the influence of an electric current and an external magnetic field applied simultaneously to a N/F point contact. Along with gaining a deeper understanding of the spin-dependent processes in such structures, we also constructed the *I–H* phase diagram of the magnetic states of a SV formed in a point contact. We expect this study to contribute to a more adequate model of SV effects observed in point contacts.

## 2. Experimental details and a model of a point contact

Figure 1 shows a schematic representation of a multi-layer thin-film structure that we used in our measurements. A Cu film with a thickness of 100 nm was deposited on a Si substrate by sputtering. This buffer layer of Cu provides a geometry in which the electric current flows nearly perpendicular to the plane of the thin film layers (CPP geometry). A ferromagnetic Co layer with a thickness of ≤ 100 nm was deposited onto the Cu layer. The Co layer was capped with a layer of either Cu or Au of several nanometers to protect it from oxidation. The surface of this thin film structure was contacted by a sharpened Cu tip using a mechanical manipulator. This tip acted as a nonmagnetic electrode, forming a point contact between the tip and the thin film structure.

The formation mechanism of the second (free) layer $F_2$ in this system is not yet clear. According to Ref. 10, apparently a very thin ferromagnetic layer $F_2$ is formed at the interface between an F-film and a nonmagnetic metallic tip N. The formation of this interface layer may be due to the inter-diffusion between the F-film and the N-tip. The magnetic properties of this interface layer $F_2$ differ from those of the bulk F-film due to a large density of nonmagnetic lattice defects and magnetic spin-lattice defects in the inter-diffusion region. Surprisingly, this layer consists of a single ferromagnetic domain, as indicated by the measured nearly-rectangular magnetoresistance hysteresis $dV/dI(H)$ loops [10]. At the boundary between the interfacial $F_2$-layer and the bulk F-film a spin-glass layer can be formed. This spin-glass layer plays the role of a spacer, which enables the magnetization of the $F_2$-layer to rotate freely with respect to the bulk F-layer. Please note that the spacer layer in this case is not a nonmagnetic metal film with relatively weak magnetic scattering, such as in conventional nano-pillar structures, but contains strongly interacting chaotically directed spins of the F-metal. Thus, to observe the SV effect in this system the thickness of this spin-glass layer must be very small, namely, of the atomic scale.

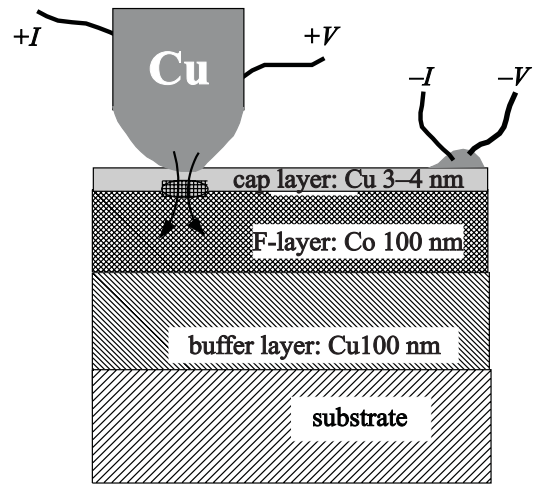

*Fig. 1*. Schematic representation of the sample layout with a buffer Cu layer, a single Co film, and a Cu or Au capping layer that is contacted by a Cu needle. Current and voltage leads that are attached to the system as shown. Small oval at the surface of the Co film represents an interfacial ferromagnetic cluster.

We would like to note that it is relatively unlikely that upon contact the inter-diffusion between the N-tip and the F-film would lead to the formation of a "free" ferromagnetic cluster surrounded by an atomically-thick spin-glass domain wall. Indeed, we observe the SV effects only in about 10% of the contacts formed between the N-tip and the F-film at random locations on the surface. As mentioned in Ref. 14, formation of magnetic clusters is possible in the interfacial region of alternately deposited Co and Cu films due to the immiscibility of the two components. A similar structure with implanted 5 nm Co clusters in a nonmagnetic layer on top of a Co film was investigated in [15]. The behavior of this special SV structure, in which the Co clusters play the role of $F_2$, is similar to what we obtain using our system. It is likely that similar clusters can be created during the formation of a mechanical contact between a tip and a thin film even at low temperatures due to metal yielding at high mechanical stress. Furthermore, since the formation of an electric contact is monitored by the onset on an electric current under an applied voltage of about 5 V, an electric discharge may accompany the contact formation. This discharge may cause "instant" melting and recrystallization of the metal in the vicinity of the contact, thus promoting the cluster formation.

This cluster should not be larger than the size of the contact in order to observe the STT effect. If the cluster is too large, then the STT effect would be too weak to change its magnetization and no hysteresis in the electric resistance would be observed. This cluster is most probably formed at the surface, since we observed the SV effects in ferromagnetic layers of only 3 nm thick and in contacts with the diameter of only 2 nm [10]. The state of such a nanostructured magnetic system differs significantly from the homogeneous state in the bulk of a ferromagnetic film. Hence, the domain wall around the cluster that is pinned at the structural inhomogeneity can be very thin, according to both the theoretical estimations [18] and experimental measurements [19,20]. The main difference of this cluster model from the "domain" model in Ref. 9 is that the size of the cluster may not exceed the size of the contact, which is of the order of several nanometers to several tens of nanometers.

Since the size of the ferromagnetic cluster should be comparable to the size of the point contact (~10 nm), the following sequential configuration of the layers is possible: bulk polycrystalline ferromagnetic film $F_1$, transitional atomically-thick spin-glass layer, and a single-domain cluster $F_2$. This ferromagnetic cluster has to be located precisely at the point contact, which explains the low probability ($\leq 10\%$) of obtaining a $F_1/N/F_2$ structure by this method. Yet the occasional observation of the SV effect in the $dV/dI(V) = R(V)$ curves and the absence of hysteresis loops in the magnetoresistance curves $dV/dI(H) = R(H)$ and *vice versa* can be explained within this model. The absence of the latter, occurring much more often than the former, can be explained if during the acquisition of the $R(H)$ curves the external magnetic field aligns the magnetization vectors of many different clusters (if these are present) up to saturation. However, the total spin torque transfer of the conductance electrons can be insufficient for the change of its magnetization direction if the size of the cluster is much larger than the diameter of the point contact. Hence, there would be no hysteresis in the $R(V)$ dependence. In a less common case, if the coercivities of the cluster and the bulk ferromagnetic layer are identical or the magnetic bond between the two is strong, the change of the magnetization direction of the cluster and the bulk layer due to an external magnetic field would proceed simultaneously. On the other hand, due to the highest current density at the point contact and, accordingly, at the cluster, the STT can lead to the change of the cluster magnetization without influencing the magnetization of the bulk layer.

The STT effect of the current on the bulk layer would be insignificant due to a substantial difference in the geometrical extension of the point contact and the layer, which results in a rapid decrease of the current density due to the spread of the current flow in the bulk of the film. In this case only a hysteresis in the $R(V)$ dependence would be observed.

### 3. Experimental results and discussion

Figure 2 shows a dependence of the differential resistance $dV/dI(V) = R(V)$ of a point contact as a function of the applied dc voltage *V*. Negative voltage (current) corresponds to a flow of electrons from the tip to the film. One can see a hysteresis of the resistance, which is caused by a change of the orientation of magnetization of the interfacial cluster (microdomain) relative to the bulk Co layer due to the STT by the electron current. The behavior of the resistance at large negative (positive) bias voltages of the order of 50–60 mV can be explained by the excitation of the magnetization vector precession in either the interfacial domain or the bulk of the film, see [21,22].

The change of the differential resistance Δ*R* in Fig. 2 constitutes 0.8%, which correlates well with the 1% height of the measured magnetoresistance curve, see the inset in Fig. 2. The slightly smaller value of Δ*R* in the $dV/dI(V)$ dependence can be attributed to an inhomogeneously distributed current density in the contact due to the lateral spread of the current. Thus, the STT effect is not sufficient for the switch of the magnetization direction in the peripheral area of the contact.

The size of the contact can be estimated by using the well-known Wexler's formula [23]:

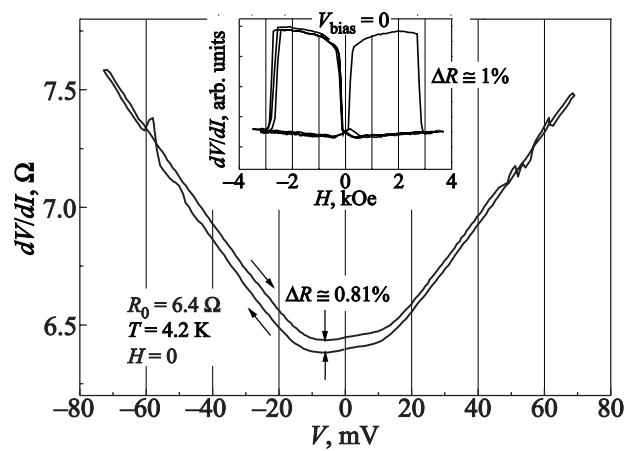

*Fig. 2.* Dependence of the differential resistance $dV/dI(V)$ of a Co–Cu point contact $R_0 = 6.4$ Ω in zero external magnetic field acquired during a bipolar current sweep. The inset shows the magnetoresistance of the same contact as a function of the applied magnetic field at zero bias voltage.

$$R_W \simeq \frac{16}{3\pi}\frac{\rho l}{d^2} + \frac{\rho}{2d}, \qquad (1)$$

where $d$ is the contact diameter, $\rho$ is the resistivity of the film material, and $l$ is the electronic mean free path. For this particular contact with the resistance of $R = 6.4\ \Omega$ we obtain $d = 20$ nm. This leads to a current density of more than $2\cdot10^9$ A·cm$^{-2}$ at a bias of $-60$ mV, at which the transition to the upper branch of the $dV/dI(V)$ curve occurs. In this estimation we used the lower limit of the resistivity $\rho = 10$ μΩ·cm for Co thin films and a value of $\rho l = 8.5\cdot10^{-12}$ μΩ·cm$^2$ (calculated for a charge carrier density of $5.8\cdot10^{22}$ cm$^{-3}$ [24]). In the calculation we neglected the resistance of the Cu electrode, which is much smaller than that of the Co film.

### 4. Magnetization reversal cycle of a surface spin valve

In the following we consider the magnetization reversal cycle of a point contact that exhibits the SV effect at different values of the electric current at positive bias voltages, see Fig. 3. A decrease of the magnetic field from its maximum value of ~ 4 kOe does not result in the change of the mutual orientation of magnetizations of the bulk layer and the surface cluster, i.e., $\mathbf{M}_1$ and $\mathbf{M}_2$, respectively. These remain parallel to the external field. The resistance of the point contact remains at its minimal value. Even at $H < 0$ coercive force keeps the positive orientation of $\mathbf{M}_1$ and $\mathbf{M}_2$ until the magnetic field reaches a value of about $-0.2$ kOe. At this field the magnetization of one of the layers flips and becomes parallel to the external field. Thus, the spin valve becomes "closed", i.e., its resistance becomes maximal, since $\mathbf{M}_1$ and $\mathbf{M}_2$ are aligned in opposite directions. Based on Refs. 13, 14, we suggest that the magnetization of the bulk film $\mathbf{M}_1$ is flipped at lower absolute values of the external magnetic field than the magnetization of the surface cluster $\mathbf{M}_2$ due to a larger coercivity of the latter. This suggestion is also supported by the absence of the influence of the bias current on the position of the slopes around $H = 0$ of the $R(H)$ curves, see Figs. 3 and 4.

If the magnetic field is swept further along the negative axis, at $H_2^- \approx -3$ kOe the magnetization of the surface clus-ter is flipped. Thus, the magnetization vectors **M**$_1$ and **M**$_2$ become parallel to each other and the resistance is reduced by $\Delta R$. Please note that the resistance decreases by the same value of $\Delta R$ as after the magnetization reversal of the bulk layer F$_1$ around $H = 0$. Usually, the resistance change $\Delta R$, which is defined as $\Delta R = (R_{AP} - R_P)/R_P$, equals to several percent, unless a multidomain structure inside the contact leads to the reduction of this value. Here $R_{AP}$ equals to the resistance of a point contact with the antiparallel magnetizations **M**$_1$ and **M**$_2$ and $R_P$ is the contact resistance at the parallel alignment of the magnetizations. The magnetization reversal of the system during a magnetic field sweep in the opposite direction occurs in a similar fashion. Thus, a complete magnetization reversal cycle of the $R(H)$ dependence consists of two symmetrically-located loops of magnetoresistance.

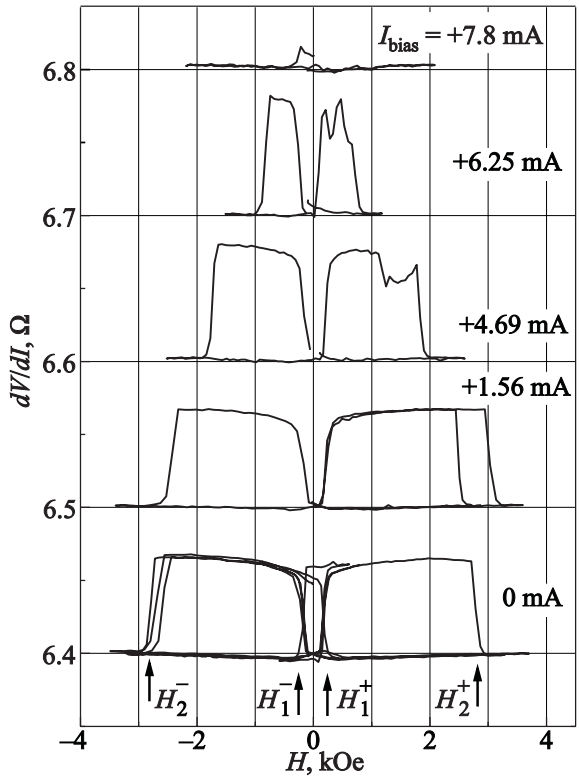

*Fig. 3.* Magnetization reversal cycles of the same contact as in Fig. 2 at different positive bias voltages (currents). At a positive bias the electrons flow from the Co film into the Cu tip. At $I = 0$ mA several curves are plotted over each other. Additionally, at this bias voltage a lesser hysteresis loop, which corresponds to the magnetization reversal of the bulk film only, was acquired during a magnetic field sweep in the region of $-0.5$ kOe $< H < +0.5$ kOe. The coercive field, which is the half-width of the small hysteresis loop, equals to about 0.17 kOe for the bulk Co layer. At 1.56 mA bias current two magnetization cycles are shown, and at 4.69, 6.25, and 7.8 mA single cycles are presented. The curves at non-zero biases are shifted along the vertical axis by 0.1 Ω for clarity. $T = 4.2$ K.

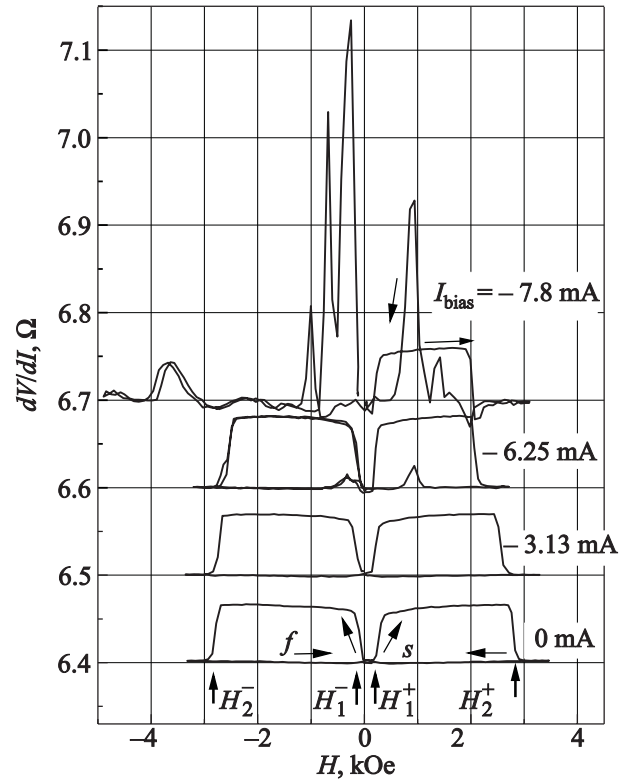

*Fig. 4.* Magnetization reversal cycles of the same contact as in Fig. 2 at different values of negative bias voltage (current). The $R(H)$ loops are observed at bias currents lower than $I = -7.8$ mA. Up to $I = -6.25$ mA the shape of the magnetization cycle curve is almost identical to the one acquired at zero bias. However, at $I = -6.25$ mA it is considerably different from the similar cycle at a positive bias of $I = +6.25$ mA in Fig. 3. Jumps in the $R(H)$ dependence are observed at a bias of $I = -7.8$ mA, which are caused by unstable magnetic states. The arrows next to the curves show the field sweep direction. *s* represents the starting point and *f* represents the ending point of the sweep in the lower curve. $T = 4.2$ K.

Magnetization cycle of the bulk layer F$_1$ is not sensitive to the changes of the polarity and the amplitude of the bias current, whereas the behavior of the surface cluster during consecutive magnetization cycles is much more susceptible to such changes, see Fig. 3. To verify that, Fig. 4 shows magnetization reversal cycles of a contact with the SV effect at negative values of the bias current.

## 5. Simultaneous influence of electric current and external magnetic field on the magnetoresistance of a point contact with the SV effect

In the following we analyze the simultaneous influence of the electric current and the external magnetic field on the magnetization direction of the interfacial single-domain cluster and, correspondingly, on the electric resistance of the whole N/F structure. A flow of spin-polarized current through the contact is accompanied by the STT effect, which influences its magnetization in addition to the external field. By analogy, we will call this influence the "STT field" in the following discussion. One can see from Fig. 3 that the magnetic field $H_2^+$ ($H_2^-$) that leads to the magnetization flipping of the cluster at zero bias is slightly less than ±3 kOe. An increase in the bias current is accompanied by a shift of the switching field $H_2$ to lower absolute values. This is the result of the cumulative action of the external and the STT fields. The latter reduces the value of the external field $H_2^\pm$ required to flip M$_2$, thus narrowing the $R(H)$-loop. At a bias current of +7.8 mA, the $R(H)$-loop disappears completely in this contact, see Fig. 3. Please note that the reversal of the magnetic field and the corresponding change of the magnetization direction of the bulk film lead to a change of the direction of the STT field. Hence, the $R(H)$ curves are virtually symmetrical relative to $H = 0$, i.e., the absolute values of $H_2^-$ and $H_2^+$ are equal. Similar influence of an electric current on the width of the $R(H)$-loops has been also observed in F$_1$/N/F$_2$ pillar structures based on Co [5].

We conclude that a positive current leads to a reduction of $H_2^\pm$. The electron flow that passes through the bulk Co layer becomes spin-polarized. Upon the entry of this flow into the surface cluster, an additional STT field is created, which results to a reduction of the width of the $R(H)$-loops.

On the other hand, if the current is negative, it helps to maintain the antiparallel alignment of **M**$_1$ and **M**$_2$, since the effective STT field now acts against the external magnetic field. As a result, one should observe an increase of the switching field $H_2^\pm$ of the surface cluster and a corresponding widening of the $R(H)$-loops. Such an effect was observed in F$_1$/N/F$_2$ structures based on permalloy (Ni$_{84}$Fe$_{16}$) [6]. However, our results show that at negative bias (current) the value of $H_2^\pm$ does not increase. Instead, it decreases with an increasing negative bias. Nevertheless, if we compare the $R(H)$-loops acquired at the same positive and negative currents (see e.g., curves at 6.25 mA in Figs. 3 and 4), we see that at negative currents the $R(H)$-loops are broader. Thus, we can conclude that there the STT field counteracts the external magnetic field.

Figure 5 shows the resulting current-field phase diagram of stable magnetic states of a surface SV. This diagram is based on the measured values of $H_2^+$ and $H_2^-$ at different values of bias voltage. Bias dependence of the coercive field $H_c$ of the surface cluster, calculated as $H_c = (H_2^+ - H_2^-)/2$ is also shown.

The absence of an increase of $H_2^\pm$ in the *I*–*H* diagram (Fig. 5) at negative currents can be linked to a spontaneous generation of magnons during the energetic relaxation of the conductance electrons, which is described theoretically in [7]. In that paper it was shown that an increase of negative bias should lead to a monotonic increase of $H_2^\pm$ in the absence of this relaxation mechanism. A slight decrease of $H_2^\pm$ observed at negative bias currents can be attributed to the influence of the magnetic (Oersted) field of the electric current and a possible heating of the contact at high DC bias. Similar effect was observed in spin valves with an interlayer of a CuPt alloy, which resulted in a current flow without spin-polarization [6].

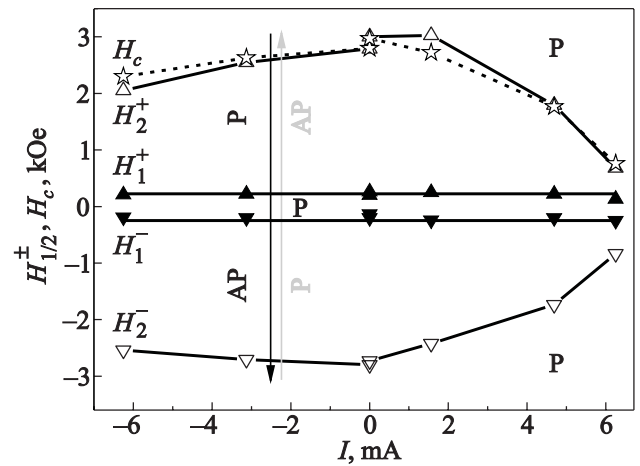

*Fig. 5.* Current-field phase diagram of magnetic states of a surface SV. The diagram is based on the measured values of $H^+$ (triangles pointing upwards) and $H^-$ (triangles pointing downwards) from the magnetoresistance curves in Figs. 3 and 4. Long arrows represent directions in the phase space during the acquisition of an $R(H)$ dependence. Stars represent the behavior of the coercive field $H_c = (H_2^+ - H_2^-)/2$ of the surface domain (cluster). Filled symbols correspond to the switching fields of the bulk layer and hollow symbols represent the switching fields of the surface cluster.

Finally, we would like to discuss the possible dynamic states of a SV. It is known that, in addition to the P–AP switching of the magnetic states of a SV, the STT effect can lead to an excitation of various vibration modes of the magnetization precession. These modes include a coherent rotation of a local magnetic moment as well as a non-coherent generation of magnons. At large currents the stationary precession of the magnetization leads to the formation of singularities, such a peaks and valleys, on the $dV/dI(V)$ dependence (see Fig. 2) due to such a dynamic behavior of the magnetization. This results in the formation of a dynamic states region on the $I–H$ phase diagram. In our measurements we have acquired $dV/dI(H)$ curves in the bias region of $V < \pm 50$ mV ($I < \pm 7$ mA, Fig. 2), i.e., we have limited ourselves to the hysteretic region of the $dV/dI(V)$ dependence where no such singularities occur.

## 6. Conclusions

The main results of this work can be summarized as follows.

1. We have proposed a model of a local surface spin valve that is formed in a N/F point contact. In this model, the bulk ferromagnetic layer acts as a spin polarizer of the electric current, which then passes through the magnetic cluster at the surface of the layer. Magnetization switching in the bulk ferromagnetic layer occurs at weaker magnetic fields than in the surface cluster. Possibly, the anisotropy of the surface magnetic cluster is larger than that of the bulk layer due to the atomic size of the cluster and its cluster-like nature. This large anisotropy results in a higher coercive field of the cluster.

2. We have constructed an $I–H$ phase diagram of stable magnetic states of a surface spin valve based on a Co–Cu microcontact. The magnetization switching field of the surface cluster decreases until its complete disappearance with an increase of the positive bias current (the spin-polarized electrons flow from the bulk film to the surface) in an in-plane external magnetic field. This effect is due to the parallel alignment of the external and the STT fields to each other leading to the enhancement of the total field. The width of the corresponding hysteresis loops in the magnetoresistance curves decreases symmetrically with an increasing negative bias. In contrast, the width of the $R(H)$-loops at negative biases is larger than their width at positive bias voltages. This effect is due to a counteraction of the STT field to the external magnetic field.

3. There is a general tendency of the $R(H)$-loops to become narrower during an increase of the absolute value of the bias independently of its polarity. This effect could be due to the reduction of the coercivity of the surface cluster by a magnetic (Oersted) field of the bias current, which can reach ~1.5 kOe for a current of $I = 7.8$ mA that flows through a contact of 20 nm diameter at $V = 50$ mV. Additionally, generation of non-equilibrium magnons and thermal effects due to a high current density of $10^9$ A/cm$^2$ could lead to the reduction of the $R(H)$-loop width.

1. S.A. Wolf, D.D. Awschalom, R.A. Buhrman, J.M. Daughton, S. von Molnar, M.L. Roukes, A.Y. Chtchelkanova, and D.M. Treger, *Science* **294**, 1488 (2001).
2. B. Dieny, V.S. Speriosu, S. Metin, S.S.P. Parkin, B.A. Gurney, P. Baumgart, and D.R. Wilhoit, *J. Appl. Phys.* **69**, 4774 (1991).
3. D.C. Ralph and M.D. Stiles, *J. Magn. Magn. Mater.* **320**, 1190 (2008).
4. M. Baibich, J.M. Broto, A. Fert, F. Nguyen Van Dau, F. Petroff, P. Etienne, G. Creuzet, A. Friederich, and J. Chazelas, *Phys. Rev. Lett.* **61**, 2472 (1988).
5. S. Urazhdin, H. Kurt, W.P. Pratt, Jr, and J. Bass, *Appl. Phys. Lett.* **83**, 114 (2003).
6. S. Urazhdin, W.P. Pratt, and J. Bass, *J. Magn. Magn. Mater.* **282**, 264 (2004).
7. S. Urazhdin, *Phys. Rev. B* **69**, 134430 (2004).
8. Y. Ji, C.L. Chien, and M.D. Stiles, *Phys. Rev. Lett.* **90**, 106601 (2003).
9. T.Y. Chen, Y. Ji, C.L. Chien, and M.D. Stiles, *Phys. Rev. Lett.* **93**, 026601 (2004).
10. I.K. Yanson, Y.G. Naidyuk, V.V. Fisun, A. Konovalenko, O.P. Balkashin, L.Y. Triputen, and V. Korenivski, *Nano Lett.* **7**, 927 (2007).
11. O.P. Balkashin, V.V. Fisun, I.K. Yanson, L.Yu. Triputen, A. Konovalenko, and V. Korenivski, *Phys. Rev. B* **79**, 092419 (2009).
12. I.K. Yanson, V.V. Fisun, Yu.G. Naidyuk, O.P. Balkashin, L.Yu. Triputen, A. Konovalenko, and V. Korenivski, *J. Phys.: Condens. Matter.* **21**, 355004 (2009).
13. I.K. Yanson, Yu.G. Naidyuk, O.P. Balkashin, V.V. Fisun, L.Yu. Triputen, S. Andersson, V. Korenivski, Y.I. Yanson, and H. Zabel, *IEEE Transactions on Magnetics* **46**, 2094 (2010).
14. M.M.H. Willekens, Th.G.S.M. Rijks, H.J.M. Swagten, and W.J.M. de Jonge, *J. Appl. Phys.* **78**, 7202 (1995).
15. X.J. Wang, H. Zou, and Y. Ji, *Appl. Phys. Lett.* **93**, 162501 (2008).
16. S.J. Steinmuller, C.A.F. Vaz, V. Ström, C. Moutafis, C.M. Gürtler, M. Kläui, J.A.C. Bland, and Z. Cui, *J. Appl. Phys.* **101**, 09D113 (2007).
17. R.A. Khan and A.S. Bhatti, *J. Magn. Magn. Mater.* **323**, 340 (2011).

18. P. Bruno, *Phys. Rev. Lett.* **83**, 2425 (1999).
19. O. Pietzsch, A. Kubetzka, M. Bode, and R. Wiesendanger, *Phys. Rev. Lett.* **84**, 5213 (2000).
20. M. Kläui, C.A.F. Vaz, J. Rothman, J.A.C. Bland, W. Wernsdorfer, G. Faini, and E. Cambril, *Phys. Rev. Lett.* **90**, 097202-2 (2003).
21. S. Urazhdin, W.L. Lim, and A. Higgins, *Phys. Rev. B* **80**, 144411 (2009).
22. S.I. Kiselev, J.C. Sankey, I.N. Krivorotov, N.C. Emley, R.J. Schoelkopf, R.A. Buhrman, and D.C. Ralph, *Nature* **425**, 380 (2003).
23. G. Wexler, *Proc. Phys. Soc*. **89**, 927 (1966).
24. W. Gil, D. Görlitz, M. Horisberger, and J. Kötzler, *Phys. Rev. B* **72**, 134401 (2005).